\documentclass[a4paper,11pt]{article}
\usepackage{pos}

\title{Equation of State of dense QCD in external magnetic field}
\ShortTitle{Equation of State of dense QCD in external magnetic field}

\author[a]{N.~Yu.~Astrakhantsev}
\author[b]{V.~V.~Braguta}
\author*[b]{N.~V.~Kolomoyets}
\author[b]{A.~Yu.~Kotov}
\author[b]{A.~A.~Roenko}

\affiliation[a]{Physik-Institut, Universit{\" a}t Z{\" u}rich, 
Winterthurerstrasse 190, CH-8057 Z{\" u}rich, Switzerland}

\affiliation[b]{Bogoliubov Laboratory of Theoretical Physics, Joint Institute for Nuclear Research, 
Dubna, 141980 Russia}

\emailAdd{nikita.astrakhantsev@physik.uzh.ch}
\emailAdd{vvbraguta@theor.jinr.ru}
\emailAdd{nkolomoyets@theor.jinr.ru}
\emailAdd{kotov.andrey.yu@gmail.com}
\emailAdd{roenko@theor.jinr.ru}

\abstract{In this proceeding we present our first results of the study of the QCD Equation of State at non-zero baryon density and in external magnetic field. We focused on the first three non-vanishing expansion coefficients of pressure in chemical potential and their dependence on magnetic field. The study is carried out within lattice simulations with $N_f=2+1$ dynamical quarks with physical quark masses. To overcome the sign problem, the simulations are carried out at imaginary baryon chemical potential. Our results suggest that external magnetic field considerably enhances the expansion coefficients and modifies their dependence on temperature. 
}

\FullConference{%
 The 38th International Symposium on Lattice Field Theory, LATTICE2021
  26th-30th July, 2021
  Zoom/Gather@Massachusetts Institute of Technology
}

\begin{document}
\maketitle

\section{Introduction}

Equation of State (EoS) of Quantum Chromodynamics (QCD) plays a fundamental role both from theoretical and practical points of view. From the theoretical perspective EoS contains an important information about thermal QCD phase transition. On the other hand, from the practical  perspective EoS is used for hydrodynamic simulations of heavy-ion collision experiments as well as in different astrophysical applications. In such applications quark-gluon matter is subject to various external conditions like high temperature, large baryon density, strong magnetic field etc. For this reason it is important to study how EoS is affected by these external conditions. 

There are a lot of phenomenological papers devoted to the calculation of the EoS under different external conditions (see, for instance, \cite{Endrodi:2013cs, Vovchenko:2017gkg, Soloveva:2021quj}). Important information about EoS was obtained by means of lattice QCD simulations. At zero baryon density it was studied in papers \cite{Borsanyi:2010cj, Borsanyi:2013bia, HotQCD:2014kol, Bernard:2006nj, Bazavov:2009zn}. Extension of lattice EoS studies to non-zero baryon chemical potential was conducted in papers \cite{Borsanyi:2012cr, Guenther:2017hnx, Bazavov:2017dus, DElia:2016jqh}. EoS for QCD in external magnetic field was studied in \cite{Bonati:2013vba, Levkova:2013qda, Bali:2014kia}. 
The second-order fluctuations of the baryon number, electric charge and strangeness, which are related to the EoS, in external magnetic field were studied in paper \cite{Ding:2021cwv}. Lattice results on the QCD phase diagram with nonzero magnetic field and baryon density can be found in~\cite{Braguta:2019yci}.  

In this Proceeding we present our first results of the study of the QCD EoS both at non-zero baryon density and in external magnetic field. 
The focus is mainly done on the expansion coefficients of pressure in chemical potential and their dependence on magnetic field.
The study is carried out within lattice simulations with $N_f = 2 + 1$ dynamical quarks with physical quark masses. To overcome the sign problem, the simulations are carried out at imaginary baryon chemical potential. 

\section{Basic definitions}

The basic quantity for the Equation of State is the pressure $p$, which can be expressed through the partition function as 
\begin{equation}
    p  = \frac{T}{V}\ln\mathcal{Z}(T, \mu_B, \mu_Q, \mu_S, eB)~,
\end{equation}
where  $V,\,T$ are spatial volume and temperature, $eB$ is external magnetic field, $\mu_B$, $\mu_Q$, $\mu_S$ are chemical potentials of the conserved baryonic, electric, and strangeness charges. The chemical potentials $\mu_B$, $\mu_Q$ and $\mu_S$ are related to the chemical potentials of individual quarks $\mu_u, \mu_d, \mu_s$ as follows
\begin{align}
  \mu_u &= \frac 1 3 \mu_B + \frac 2 3 \mu_{Q}~, \nonumber \\
  \mu_d &= \frac 1 3 \mu_B - \frac 1 3 \mu_{Q}~, \\
    \mu_s &= \frac 1 3 \mu_B - \frac 1 3 \mu_{Q} - \mu_S~. \nonumber
\end{align}
In our first exploratory study we consider a simple particular combination of chemical potentials:
\begin{equation}
\mu_u=\mu_d=\mu\,,\qquad \mu_s=0~,
\end{equation}
what implies $\mu_B=3 \mu,\, \mu_Q=0,\, \mu_S=\mu$. For this parameterization and for sufficiently small chemical potential $\mu$, the EoS can be expanded in powers of $\theta = \mu/T$. We restrict our consideration by the first four non-zero terms in this expansion
\begin{equation}
    \frac {p} {T^4} = c_0 + c_2 \theta^2 + c_4 \theta^4 + c_6 \theta^6~. 
\end{equation}
It is important to notice that the coefficients $c_2, c_4, c_6$ are related to the fluctuations and correlations of the conserved charges. In particular, $c_2$ can be represented in the following way
\begin{align}
 c_2 &= \frac 1 {2} (9 \chi_2^B + 6 \chi_{11}^{BS} + \chi_2^S )~,
\label{eq::c2_from_chi}
\end{align}
where we used the designations 
\begin{equation}
 \chi_{ijk}^{BQS} = \frac {\partial^{i+j+k} p/T^4} {\partial (\mu_B/T)^i \partial (\mu_Q/T)^j \partial (\mu_S/T)^k }~.
\end{equation}
In this Proceeding we focus on the coefficients $c_2, c_4, c_6$ and their dependence on external magnetic field. 

\section{Lattice setup}

In our study we consider the partition function for
${N_f=2+1}$ QCD with chemical potentials $\mu_f$ ($f = u,d,s$) coupled
to quark number operators, ${\mathcal Z}(T,\mu_u,\mu_d,\mu_s,eB)$, in a
setup 
$\mu_u = \mu_d =\mu$, $\mu_s=0$.  The path integral formulation of 
${\mathcal Z} (T,\mu_B,eB)$, discretized using improved rooted staggered
fermions and the standard exponentiated implementation of the
chemical potentials, reads
\begin{equation}
  \mathcal{Z} = \int \mathcal{D}U e^{- \mathcal{S}_{\text{YM}}}
  \prod_{f=u,d,s}\det\left[ M_{\text{st}}^{f}(U,\mu_{f}) \right]^{1/4} ~, 
\label{partfunc}
\end{equation}
where
\begin{equation}
  \mathcal{S}_{\text{YM}} = -\frac{\beta}{3} \sum_{i,\mu\neq\nu}
  \left( \frac{5}{6}W_{i;\mu\nu}^{1\times1}
    - \frac{1}{12}W_{i;\mu\nu}^{1\times2} \right)
\end{equation}
is the tree-level Symanzik improved action
($W_{i;\mu\nu}^{n\times m}$ stands for the trace of
the $n\times m$ rectangular parallel transport in the $\mu$-$\nu$
plane and starting from site $i$), and the staggered fermion matrix is
defined as
\begin{equation}
  M_{\text{st}}^{f}(U,\mu_{f})
   =
        am_f\delta_{i,j} +
        \sum_{\nu=1}^{4} \frac{\eta_{i;\nu}}{2}\big[
        e^{ a\mu_{f}\delta_{\nu,4}} u^f_{i;\,\nu} U_{i;\nu}^{(2)}
        \delta_{i,j-\hat{\nu}}
  -  e^{-a\mu_{f}\delta_{\nu,4}} u^{f*}_{i-\hat\nu;\,\nu} U_{i-\hat{\nu};\nu}^{(2)\dagger}
        \delta_{i,j+\hat{\nu}}
        \big]~,
  \label{fermatrix}
\end{equation}
where $U_{i;\nu}^{(2)}$ are two-times stout-smeared links, with
isotropic smearing parameter $\rho=0.15$~\cite{Morningstar:2003gk} and $u^f_{i;\,\mu}$ is the Abelian field phase.
The Abelian transporters corresponding to a uniform magnetic field $B_z$ directed along $\hat{z}$ axis are chosen in the standard way leading to the quantization condition
\begin{equation}\label{bquant}
\frac{e}{3}B_z=\frac{2 \pi b}{a^2 N_x N_y}\ ,
\end{equation}
where $b$ is an integer.

Bare parameters have been set so as to stay on a line of constant
physics~\cite{Aoki:2009sc, Borsanyi:2010cj}, with
equal light quark masses, $m_u=m_d=m_l$, {a physical
strange-to-light mass
ratio, $m_s/m_l=28.15$, and a physical pseudo-Goldstone pion mass, 
$m_{\pi}\simeq135~\text{MeV}$.}

The simulations were carried out on the lattice $6\times 24^3$ 
for $eB=0,~0.5,~0.6,~0.8,~1.0,~1.5$~GeV$^2$ for a set of temperatures and chemical potentials $\mu$. In addition we conducted 
simulations on the lattices $8 \times 32^3$ and $10 \times 40^3$ for $eB=0.6$~GeV$^2$ for a set of temperatures and chemical potentials $\mu$. We used $O(100)$ statistically independent configurations for each set of lattice parameters used in our study. 

One cannot measure directly the partition function and pressure in lattice simulations. Instead of it we measured the quark number density $n$ and determined the coefficients $c_2, c_4, c_6$ of the expansion:
\begin{equation}
\frac {n} {T^3}= \frac {\partial p/T^4} {\partial \theta} = 2 c_2 \theta + 4 c_4 \theta^3 +6 c_6 \theta^5.
\label{density}
\end{equation}

\section{The results of the calculations}

\begin{figure}
    \centering
    \includegraphics[width = \textwidth]{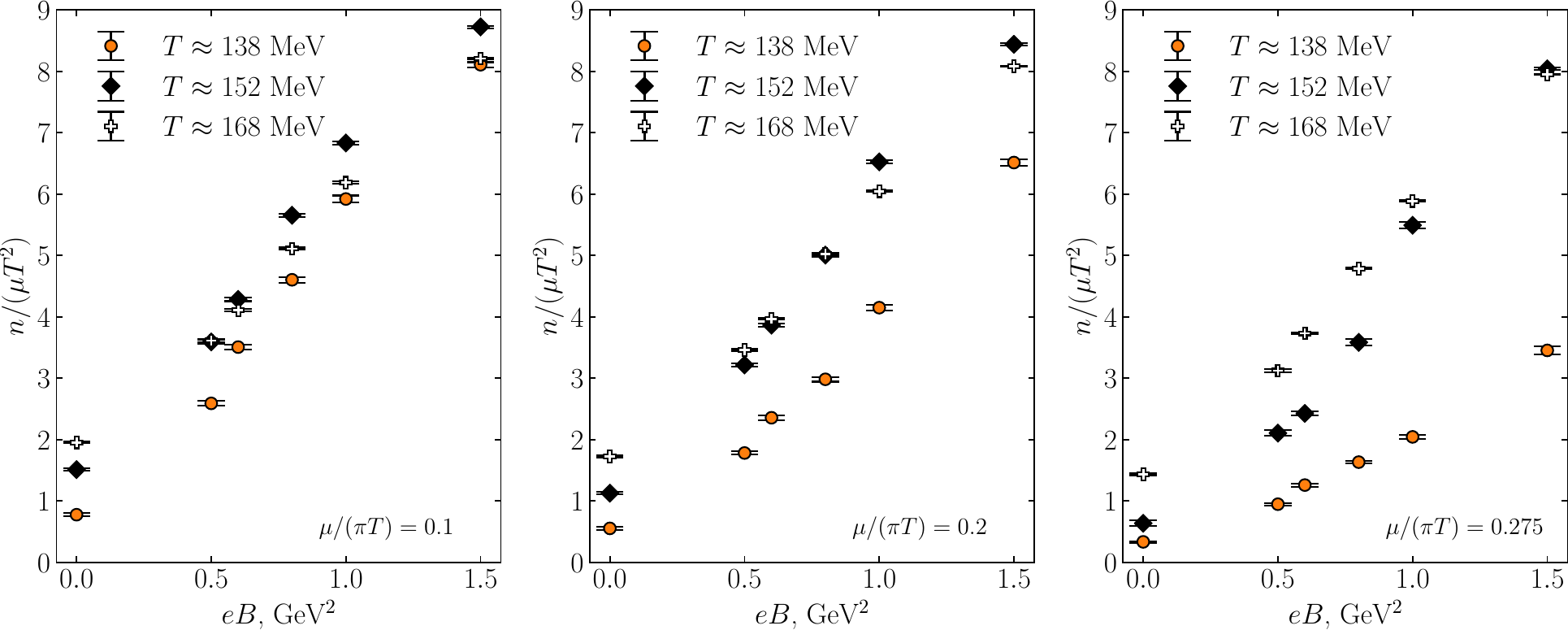}
    \caption{The ratio $n /\mu T^2$ as a function of magnetic field for various values of the chemical potential $\mu$ and temperature.}
    \label{fig:nOverMuT2-vs-eB}
\end{figure}

In Fig.~\ref{fig:nOverMuT2-vs-eB} we show the ratio $n /\mu T^2$ as a function of magnetic field for various values of the chemical potential $\mu$ and temperature. 
Fitting the data for the density $n$ by formula~(\ref{density}) we determine the coefficients $c_2, c_4, c_6$. 

\begin{figure}
    \centering
    \includegraphics[width = \textwidth]{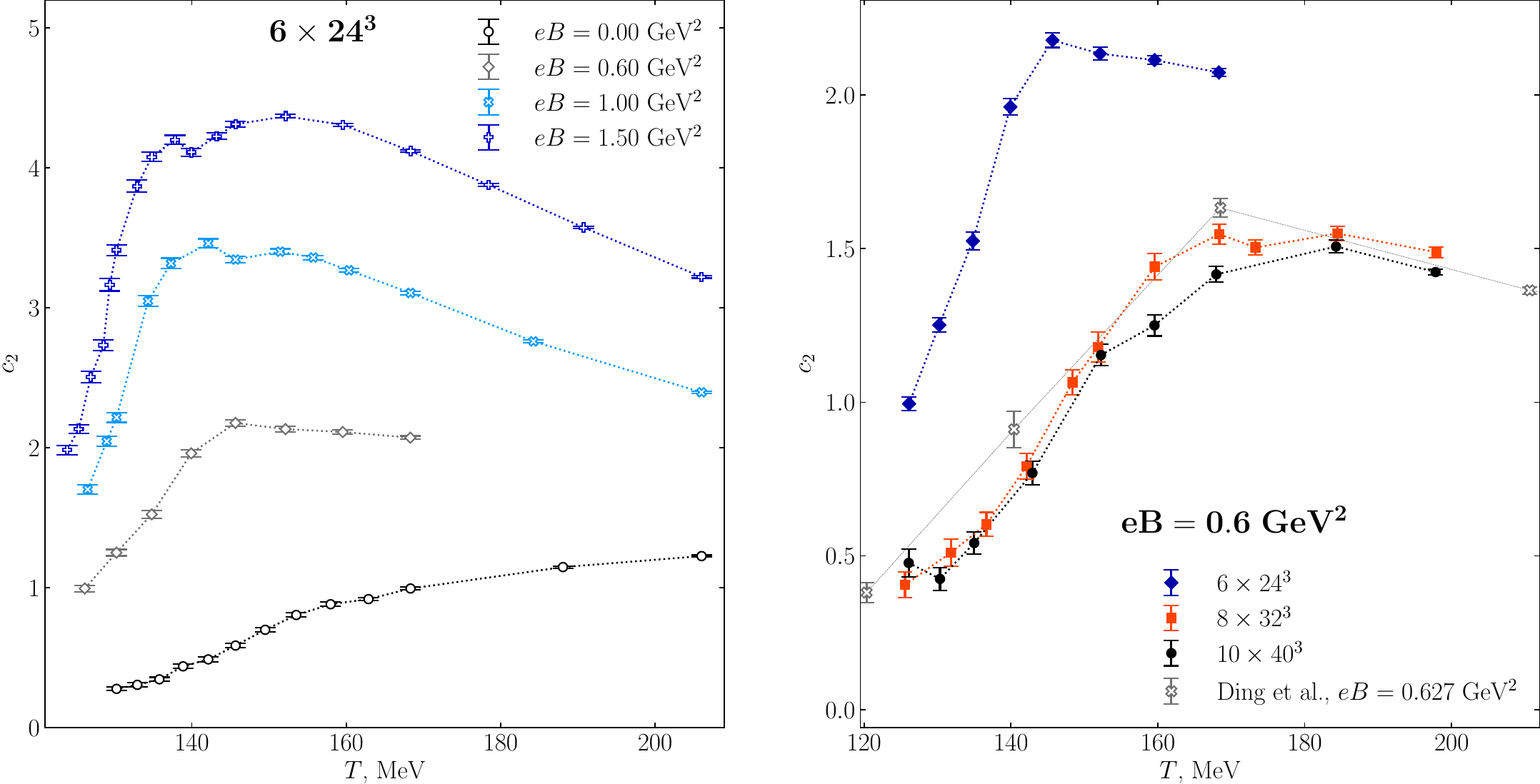}
    \caption{
    On the left panel: the coefficient $c_2$ as a function of temperature for various magnetic fields on the lattice $6\times24^3$. On the right panel: the $c_2$ coefficient for the field $eB=0.6$\,GeV$^2$ and the lattices $6\times24^3,\,8\times32^3,\,10\times40^3$. In addition, we plot the $c_2$ coefficient determined from the results of paper \cite{Ding:2021cwv} at $eB=0.627$\,GeV$^2$. Lines are to guide the eye.
    }
    \label{fig:c2withDing}
\end{figure}

On the left panel of Fig.~\ref{fig:c2withDing}  we plot the coefficient $c_2$ as a function of temperature for various magnetic fields. These results were obtained on the lattice $6\times24^3$.  
One sees that magnetic field considerably enhances the value of the coefficient $c_2$, i.e. the fluctuations. Notice also that the phase transitions in QCD manifest themselves as an inflection point of the $c_2$. However, at sufficiently strong magnetic field this inflection point turns into a peak which shifts to the lower temperatures at larger magnetic fields. We believe that the behaviour of the peak position can be associated with the decrease of the critical temperature by magnetic field~\cite{Bruckmann:2013oba}. The height of the peak also increases which implies that magnetic field enhances fluctuations at the QCD phase transition point. These properties of the $c_2$ coefficient which we observed in our study are in agreement with that observed in paper~\cite{Ding:2021cwv}. Notice, that in \cite{Braguta:2019yci} we also observed a change of dense QCD properties at similar values of magnetic field $eB=eB^{\text{fl}}\sim 0.6$ GeV$^2$, in particular, the dependence of the width of the chiral thermal phase transition on the value of chemical potential changed direction at $eB^{\text{fl}}$.

To estimate the discretization effects, the comparison of results obtained on different lattices is done. On the right panel of Fig.\,\ref{fig:c2withDing}, the $c_2$ coefficient obtained from $6\times24^3$, $8\times32^3$ and $10\times40^3$ lattices is shown. One can see that the discretization effects obtained on the smallest lattice are quite large. At the same time, the coefficients calculated on $8\times32^3$ and $10\times40^3$ lattices are in agreement with each other. Moreover, the results obtained on two larger lattices agree with ones of paper \cite{Ding:2021cwv}. The last are determined from results of that investigation for $eB = 0.627\text{\,GeV}^2$ using~Eq.\,(\ref{eq::c2_from_chi}).

\begin{figure}
    \centering
    \includegraphics[width = \textwidth]{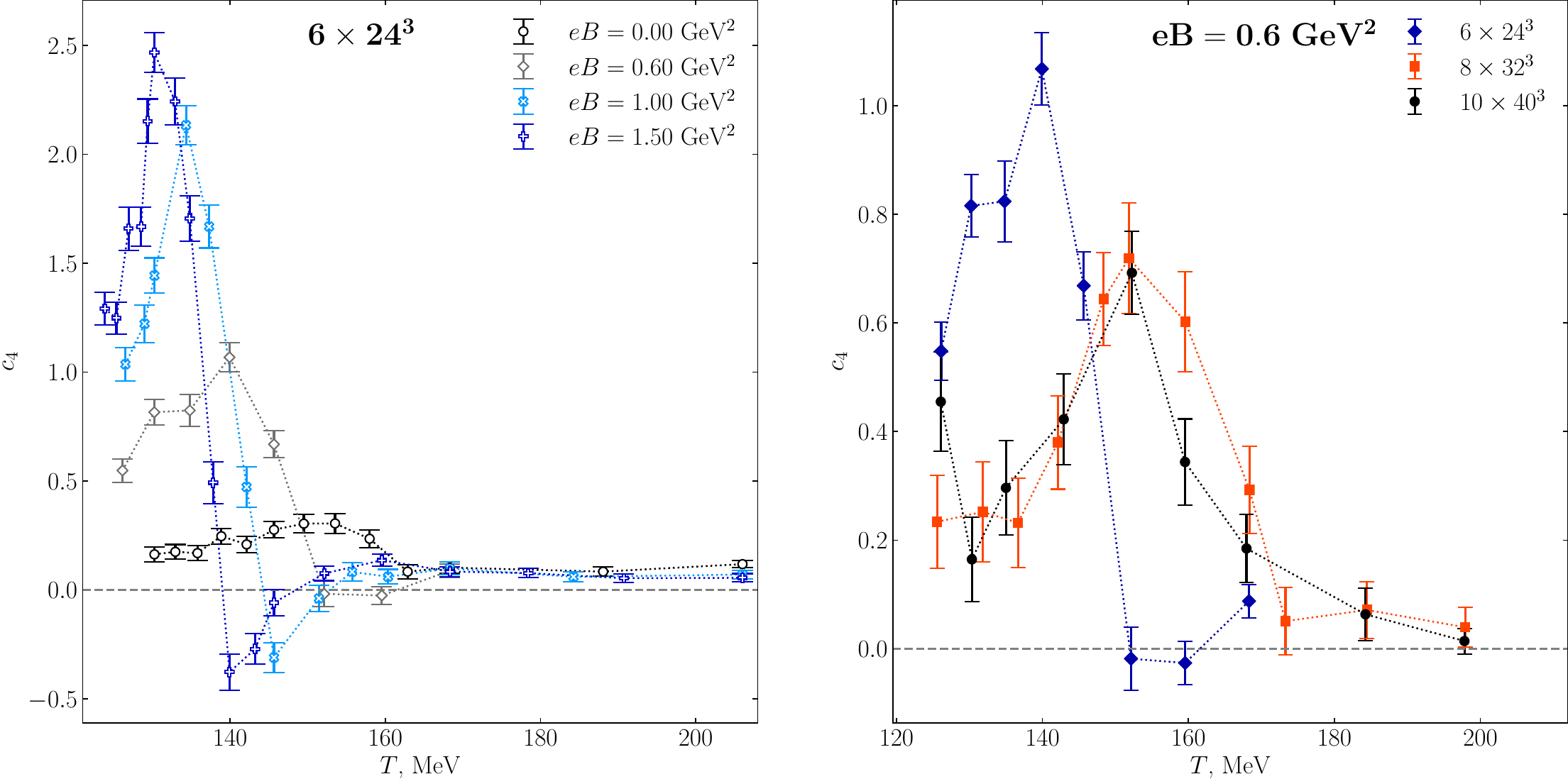}
    \caption{ On the left panel: the coefficient $c_4$ as a function of temperature for various magnetic fields on the lattice $6\times24^3$. On the right panel: the $c_4$ coefficient for the field $eB=0.6$\,GeV$^2$ and the lattices $6\times24^3,\,8\times32^3,\,10\times40^3$. Lines are to guide the eye.}
    \label{fig:c4_3Lat}
\end{figure}

\begin{figure}
    \centering
    \includegraphics[width = \textwidth]{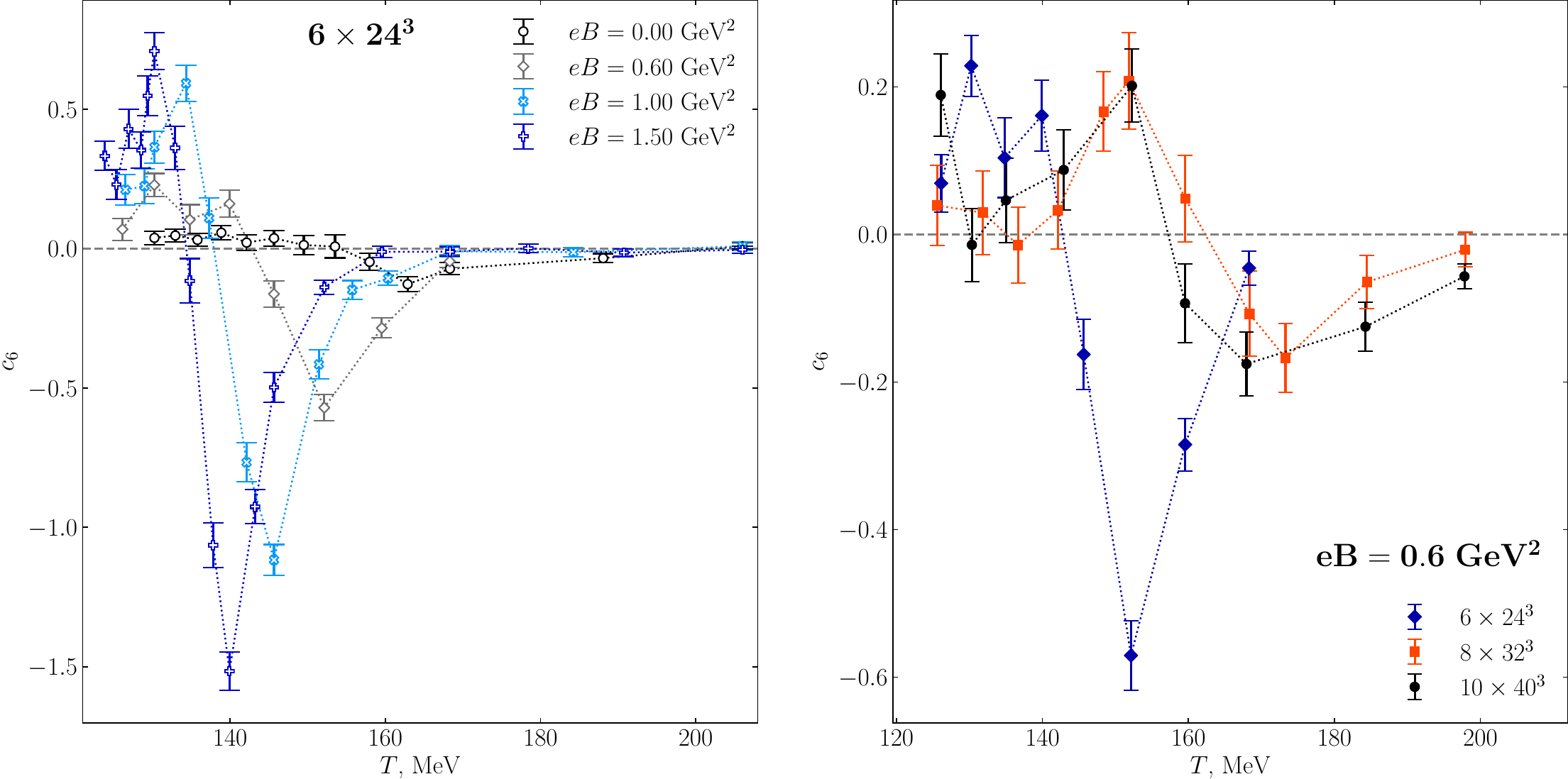}
    \caption{ On the left panel: the coefficient $c_6$ as a function of temperature for various magnetic fields on the lattice $6\times24^3$. On the right panel: the $c_6$ coefficient for the field $eB=0.6$\,GeV$^2$ and the lattices $6\times24^3,\,8\times32^3,\,10\times40^3$. Lines are to guide the eye.}
    \label{fig:c6_3Lat}
\end{figure}

In Fig.\,\ref{fig:c4_3Lat} and Fig.\,\ref{fig:c6_3Lat} we plot the coefficients $c_4$ and $c_6$ as a function of temperature for various magnetic fields under consideration. The results obtained on the lattice $6\times24^3$ are shown on the left panels of these figures. 
The comparison of the results calculated on different lattices at $eB=0.6$~GeV$^2$ is shown on the right panel of Fig.\,\ref{fig:c4_3Lat} and Fig.\,\ref{fig:c6_3Lat}. Similarly to the $c_2$ coefficient, the $c_4$ and $c_6$ 
coefficients are considerably enhanced by magnetic field. Moreover, at sufficiently large magnetic field the coefficients $c_4$ and $c_6$ change their behaviour with temperature. They start to flip sign at some temperature. Notice, however, as can be seen from right panel of Fig.\,\ref{fig:c4_3Lat} and Fig.\,\ref{fig:c6_3Lat}, the discretization effects are quite large for lattice size $6\times24^3$, while for larger lattice sizes $8\times32^3$, $10\times40^3$ data exhibit very mild dependence on the lattice size.  

\section{Discussion and conclusion}

In this Proceeding we presented our first results on the study of the QCD EoS at non-zero baryon density and in external magnetic field. We focused on the three non-vanishing expansion coefficients of pressure in chemical potential and their dependence on magnetic field. The study is carried out within lattice simulations with $N_f=2+1$ dynamical quarks with physical quark masses. To overcome the sign problem, the simulations are carried out at imaginary baryon chemical potential. In our study
we found that external magnetic field considerably enhances the expansion coefficients and modifies their temperature dependence. We observe, that at large magnetic fields the coefficient $c_2$ exhibits a peak in temperature dependence and coefficients $c_4$ and $c_6$ change sign at some temperature.

Comparing the values of the coefficients obtained on different lattices one can state that our results contain noticeable systematic uncertainty. We are going to reduce these uncertainties in the forthcoming study. 

Despite the systematic uncertainty in the results, we believe that the main our conclusion remains to be true. The expansion coefficients of the EoS strongly depend on magnetic field. This dependence might be explained by asymmetry between parallel and perpendicular to the external magnetic field directions~\cite{Astrakhantsev:2019zkr}, which effectively reduces dimension of the system under study. 

\acknowledgments
The authors are grateful to Alexander Nikolaev for the help with the numerical calculations.
This work was supported by RFBR grant 18-02-40126.
This work has been carried out using computing resources of the Federal collective usage center Complex for Simulation and Data Processing for Mega-science Facilities at NRC ``Kurchatov Institute'', \url{http://ckp.nrcki.ru/}; and the Supercomputer  ``Govorun'' of Joint Institute for Nuclear Research. 

\bibliographystyle{JHEP}

\bibliography{biblio/biblio.bib}

\end{document}